\documentclass[aps,prl,showpacs,twocolumn]{revtex4-1}

\usepackage[english]{babel}
\usepackage[utf8]{inputenc}
\usepackage{amsmath}
\usepackage{amssymb}
\usepackage[caption=false]{subfig}
\usepackage{amssymb}
\usepackage{epsfig}
\usepackage{graphicx}
\usepackage{amsmath}
\usepackage{array,color}
\usepackage{natbib}

\usepackage[usenames,dvipsnames]{xcolor}
\definecolor{forestgreen}{rgb}{0.11,0.54,0.15}
\definecolor{purple}{rgb}{0.62,0.10,0.96}
\definecolor{dockerblue}{rgb}{0.11,0.56,0.98}
\definecolor{freeblue}{rgb}{0.25,0.41,0.88}

\usepackage[pdftex,plainpages=false,colorlinks=true,linkcolor=Red, citecolor=blue, urlcolor=blue]{hyperref}

\begin{document}

\title{Maximum Entropy Analytic Continuation for Spectral Functions \\with Non-Positive Spectral Weight}
\author{A. Reymbaut$^1$, D. Bergeron$^1$ and A.-M. S. Tremblay$^{1,2}$}
\affiliation{
$^1$D\'{e}partement de physique and Regroupement qu\'{e}b\'{e}cois sur les mat\'{e}riaux de pointe, Universit\'{e} de Sherbrooke, Sherbrooke, Qu\'{e}bec, Canada J1K 2R1 \\
$^2$Canadian Institute for Advanced Research, Toronto, Ontario, Canada, M5G 1Z8
}
\date{\today}
\begin{abstract}
Information about the pairing mechanism for superconductivity is contained in the spectral weight for the anomalous (Gorkov) Green function. In the most general case, this spectral weight can change sign on the positive real axis or even be complex in the presence of broken time-reversal symmetry. This renders impossible the direct analytic continuation with Maximum Entropy methods of numerical results obtained for the anomalous Green function either in Matsubara frequency or in imaginary time. Here we show that one can define auxiliary Green functions that allow one to extract the anomalous spectral weight in the most general multi-orbital spin-singlet or triplet cases with no particular symmetry. As a simple example, we treat the case of lead in Eliashberg theory: Maximum Entropy analytic continuation with auxiliary Green functions (MaxEntAux) allows us to recover spectral functions that change sign on the positive real axis. The approach can be extended to transport quantities.
\end{abstract}
\pacs{74.20.-z, 74.20.Mn}
\maketitle



Not so long after the proposal of the phonon mechanism for superconductivity in BCS theory \cite{BCS1,BCS:1957}, the central role played by retardation was recognized \cite{Morel:1962,Eliashberg:1960}. The convincing proofs of the validity of the phonon pairing mechanism, implicit in BCS theory \footnote{For a review of conventional superconductivity that includes a discussion of historical matters, see the review by Carbotte and Marsiglio in Ref.~\protect\cite{MarsiglioCarbotteBook:2007} and in the book by Parks~\protect\cite{Parks:1969}}, came from small deviations from BCS explained by the more refined theory of Migdal and Eliashberg ~\cite{Migdal:1958,Eliashberg:1960}. One of the most direct proofs came through the frequency dependence of the gap function measured through tunneling experiments \cite{Rowell:1963}. Indeed, signatures of phonon frequencies and their harmonics observed in the frequency dependence of the density of states were explained in detail by theory \cite{Scalapino:1963,McMillan:1965,Scalapino:1966}.

In the search for pairing mechanisms in unconventional superconductors, such as cuprates and organic superconductors, one must face the presence of strongly-correlated physics, manifest through the proximity between the Mott insulating and superconducting phases. The discussion between P. W. Anderson \cite{Anderson:2007}, and D. J. Scalapino \cite{Scalapino_E-letter, Scalapino_RMP:2012} shows that strongly-correlated physics sets again the question of retardation, or more generally the frequency dependence of pairing, at the heart of the problem. New high-energy pairing mechanisms could be involved.  

The thorough theoretical study of this subject is not an easy one as it requires to have access to the frequency dependent pairing dynamics embodied in the anomalous self-energy or in the Gorkov function \cite{Gorkov:1958}. 
While direct real-frequency studies are sometimes possible \cite{Maier:2008,Kyung:2009,SenechalResilience:2013,Sakai_Civelli_Imada_arxiv:2014},  most calculations, especially at finite temperature, rely on Matsubara-frequency or imaginary-time calculations. Direct analytic  continuation with Pad\'e approximants is occasionally possible \cite{Serene:1977,Beach:2000}, but for Quantum Monte Carlo data in particular \cite{Gull_Millis_2014_pairing_glue, Gull_Millis_2015_pairing_glue}, one must rely on Maximum Entropy methodology for analytic  continuation to the real axis \cite{Jarrell:1996}. However, this method works only for spectral functions whose sign does not change over the positive or negative frequency range. The anomalous self-energy and spectral function for the Gorkov function do not satisfy this criterion. A workaround using a constant shift has been suggested \cite{JarrellJulichPavarini:2012}, but this involves another adjustable parameter and is not valid for complex spectral functions. 

We first consider the simple case of singlet pairing in a one-band model and give an example of application with the gap function for lead where the answer is known. The more general case is considered at the end. 

\paragraph{Auxiliary Green Function for Analytic Continuation of $\mathcal{F}$} 

Green functions associating a creation or annihilation operator to its hermitian conjugate do have a constant sign spectral weight and can be analytically continued using Maximum Entropy methodology. With this in mind, following Ref.\cite{Gull_Millis_2014_pairing_glue}, define the mixed operator
\begin{equation}
\hat{a}_{\vec{k}} = \hat{c}_{\vec{k}\uparrow} + \hat{c}^\dagger_{-\vec{k}\downarrow},
\label{Eq_mixed_operator}
\end{equation}
and the auxiliary Green function 
\begin{equation}
\mathcal{G}^{aux}(\vec{k},\tau) = -\left\langle \hat{\mathcal{T}}_{\tau}\, \hat{a}_{\vec{k}}(\tau)\, \hat{a}^\dagger_{\vec{k}}(0) \right\rangle_{\hat{\mathcal{H}}}.
\label{Eq_Total_Green_Function_Simple}
\end{equation}
where the average is with respect to the grand canonical ensemble for Hamiltonian $\hat{\mathcal{H}}$ and the imaginary-time ordering operator is $\hat{\mathcal{T}_{\tau}}$. In terms of the original annihilation (creation) operators $\hat{c}^{(\dagger)}$, the auxiliary Green function is given by
\begin{eqnarray}
\mathcal{G}^{aux}(\vec{k},\tau)& =&
 -\left\langle \hat{\mathcal{T}}_{\tau}\, \hat{c}_{\vec{k}\uparrow}(\tau)\, \hat{c}^\dagger_{\vec{k}\uparrow}(0) \right\rangle_{\hat{\mathcal{H}}}
-\left\langle \hat{\mathcal{T}}_{\tau}\, \hat{c}^\dagger_{-\vec{k}\downarrow}(\tau)\, \hat{c}_{-\vec{k}\downarrow}(0) \right\rangle_{\hat{\mathcal{H}}}\nonumber\\
&-&\left\langle \hat{\mathcal{T}}_{\tau}\, \hat{c}_{\vec{k}\uparrow}(\tau)\, \hat{c}_{-\vec{k}\downarrow}(0) \right\rangle_{\hat{\mathcal{H}}}
-\left\langle \hat{\mathcal{T}}_{\tau}\, \hat{c}^\dagger_{-\vec{k}\downarrow}(\tau)\, \hat{c}^\dagger_{\vec{k}\uparrow}(0) \right\rangle_{\hat{\mathcal{H}}}.
\label{Gaux}
\end{eqnarray}
If the maximum-entropy analytic continuation is done from imaginary time, it suffices to analytically continue both $\mathcal{G}^{aux}(\vec{k},\tau)$  and  $\mathcal{G}_{\uparrow\downarrow}(\pm\vec{k},\tau)=-\left\langle \hat{\mathcal{T}}_{\tau}\, \hat{c}_{\pm\vec{k}{\uparrow\downarrow}}(\tau)\hat{c}^\dagger_{\pm\vec{k}{\uparrow\downarrow}}(0) 
\right\rangle_{\hat{\mathcal{H}}}$ that have positive spectral weight and to extract the spectral weight for the off-diagonal terms (Gorkov functions) by taking the difference. This is possible because in the case considered, the Gorkov function $\mathcal{F}(\vec{k},\tau)=-\left\langle \hat{\mathcal{T}}_{\tau}\, \hat{c}_{\vec{k}\uparrow}(\tau)\, \hat{c}_{-\vec{k}\downarrow}(0) \right\rangle_{\hat{\mathcal{H}}}$ and its complex conjugate have identical spectral weight. Indeed, for singlet pairing with time-reversal symmetry, the spectral weight $\mathcal{A}^{an}(\vec{k},\omega)$ is real and odd in frequency, implying that the Gorkov function in Matsubara frequency is also real 
\begin{eqnarray}
\mathcal{F}(\vec{k},i\omega_n) & = & \int \!\frac{\mathrm{d}\omega}{2\pi} \,\frac{\mathcal{A}^{an}(\vec{k},\omega)}{i\omega_n -\omega} \nonumber \\
& = & -\int \!\frac{\mathrm{d}\omega}{2\pi} \,\frac{\omega\,\mathcal{A}^{an}(\vec{k},\omega)}{\omega_n^2 +\omega^2} \in \mathbb{R}.
\end{eqnarray}
Hence, the Matsubara frequency result
\begin{equation}
\mathcal{G}^{aux}(\vec{k},i\omega_n) = \mathcal{G}_{\uparrow} (\vec{k},i\omega_n) - \mathcal{G}_{\downarrow} (\vec{k},-i\omega_n) +2\mathcal{F}(\vec{k},i\omega_n)
\label{Eq_Trick_iwn}
\end{equation}
immediately translates into
\begin{equation}
\mathcal{A}^{an}(\vec{k},\omega) = \frac{1}{2}\, \left[ \mathcal{A}^{aux}(\vec{k},\omega) - \mathcal{A}_{\uparrow}(\vec{k},\omega) - \mathcal{A}_{\downarrow}(\vec{k},-\omega) \right]. 
\label{Eq_Trick}
\end{equation}
with no restriction to the particle-hole symmetric case, as assumed in Ref.\cite{Gull_Millis_2014_pairing_glue}. The spectral weight for the auxiliary Green function  $\mathcal{A}^{aux}(\vec{k},\omega)$ is normalized to two instead of unity. 


\paragraph{An Example with Sign Changing Gap Function}

Let us focus on the analytic continuation of the solutions to the Matsubara frequency Eliashberg equations for lead. Usually, this analytic continuation is performed with Pad\'e approximants. Here, we take published real-frequency results~\cite{Scalapino:1966} that we transform to Matsubara frequency. Our method, MaxEntAux, is then tested by analytically continuing these to real frequencies and comparing with the original data. 

\begin{figure}[ht!]
	\begin{center}
		\includegraphics[width=0.48\textwidth]{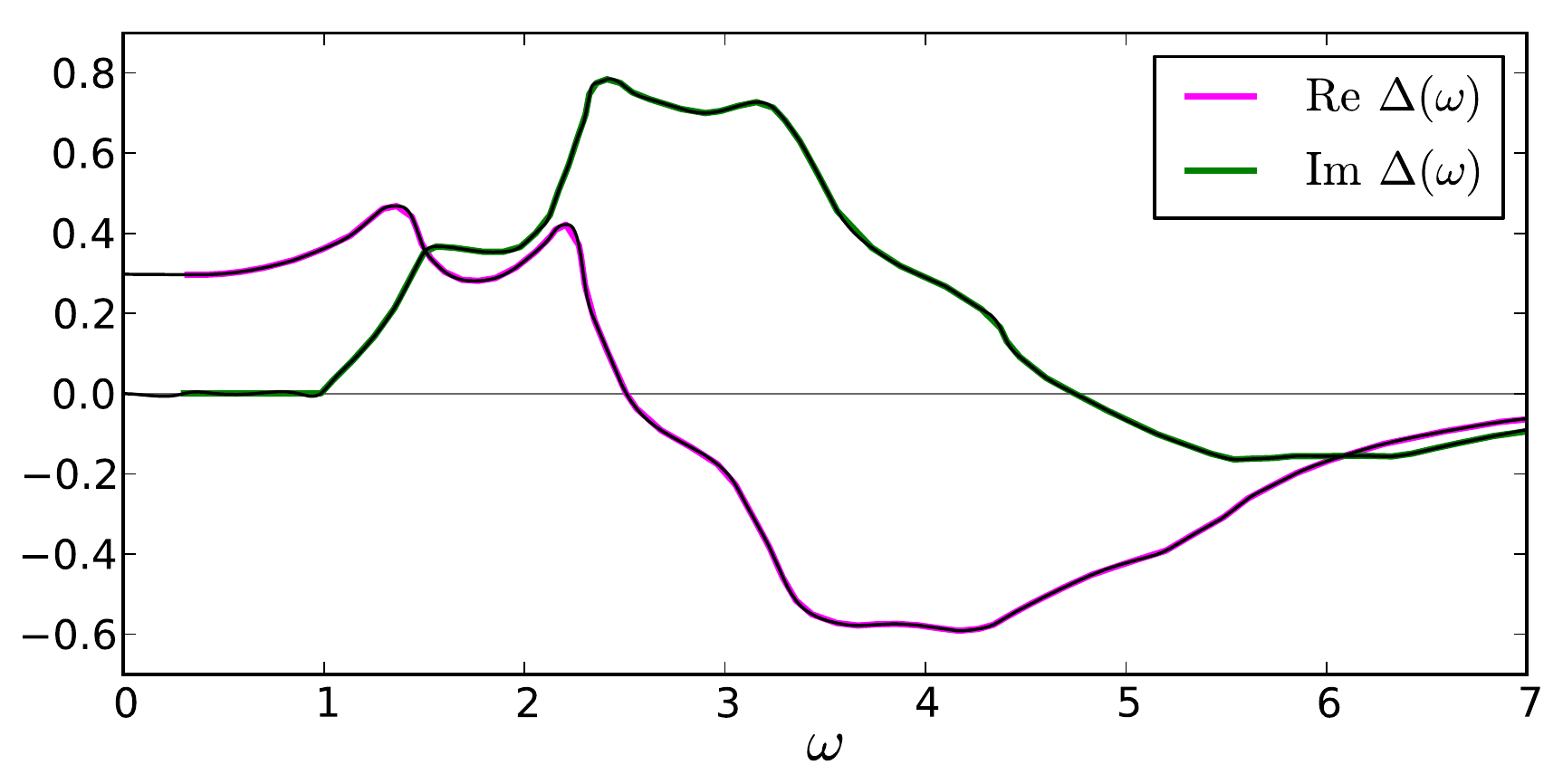}
		\caption{Solid magenta and green lines : real and imaginary parts of the gap function extracted from Ref.\cite{Scalapino:1966}. Black solid lines : spline interpolations of the data. Note that these interpolations are extrapolated within the gap with a polynomial function that satisfies the continuity of the gap function and its derivative, and has a zero slope at $\omega=0$. ${\rm Im \,\Delta}$ is odd in $\omega$ and ${\rm Re\, \Delta}$ even.}
		\label{Fig_Interpolations_Delta}
	\end{center}
\end{figure}

Consider the spline interpolation $\Delta(\omega)$ of the gap function extracted from Ref.\cite{Scalapino:1966}, shown in Fig.\ref{Fig_Interpolations_Delta}. The transverse phonon frequency $\omega_1^t = 4.4$ meV is taken as unity. From this gap function, one can work out the local and anomalous spectral functions
\begin{eqnarray}
\mathcal{A}^{loc}(\omega) & = & \mathcal{C} \; \mathrm{Re}\left( \frac{\omega}{\sqrt{\omega^2 - \Delta^2(\omega)}} \right)  \label{Eq_DOS}\\
\mathcal{A}^{an}(\omega) & = & \mathcal{C}\;  \mathrm{Re}\left( \frac{\Delta(\omega)}{\sqrt{\omega^2 - \Delta^2(\omega)}} \right) .
\label{Eq_DOS_SC}
\end{eqnarray}
where the square root is chosen such that the real part is odd in $\omega$ and the imaginary part even, $\mathcal{C}$ is the constant normalizing $\mathcal{A}^{loc}(\omega)$ to unity over the frequency interval $[-\omega_c,\omega_c]$, where the cutoff frequency is $\omega_c = 25$. Note that since the published real and imaginary parts of $\Delta(\omega)$ extend only up to $\omega\sim 7$, the resulting spectral functions \eqref{Eq_DOS} and \eqref{Eq_DOS_SC} are extended as constants up to $\omega_c $ such that both real and imaginary parts of $\Delta(\omega)$ obtained from Kramers-Kr\"onig relations and from Eq.\eqref{Eq_Delta} agree with the published results~\cite{Scalapino:1966}.

Using the spectral representations 
\begin{eqnarray}
\mathcal{G}^{loc}(i\omega_n) & = & \int \!\frac{\mathrm{d}\omega}{2\pi}\, \frac{\mathcal{A}^{loc}(\omega)}{i\omega_n - \omega} \\
\mathcal{F}(i\omega_n) & = & \int \!\frac{\mathrm{d}\omega}{2\pi}\, \frac{\mathcal{A}^{an}(\omega)}{i\omega_n - \omega},
\end{eqnarray}
Eqs.~\eqref{Eq_DOS} and \eqref{Eq_DOS_SC} give the local and anomalous Green functions in Matsubara frequency for  $T=1.3~{\rm K}=0.02~\omega_1^t$.  The integral over wave vectors of Eq.\eqref{Eq_Trick_iwn} yields
\begin{equation}
\mathcal{G}^{aux}(i\omega_n) = \mathcal{G}^{loc} (i\omega_n) - \mathcal{G}^{loc} (-i\omega_n) + 2 \mathcal{F}(i\omega_n).
\label{Eq_aux_Matsubara}
\end{equation}

We add a gaussian noise to impose a relative error of $10^{-4}$ on each Matsubara frequency component of the functions in Eq.\eqref{Eq_aux_Matsubara} to simulate Monte Carlo statistical errors.~\footnote{Even with a relative error of $10^{-3}$, the sign change in analytically continued functions is very clear. The low frequency cumulative order parameter in particular remains reliable.} Using the Maximum Entropy code OmegaMaxEnt~\cite{Bergeron:2015}, we then analytically continue $\mathcal{G}^{loc}(i\omega_n)$ and $\mathcal{G}^{aux}(i\omega_n)$ to real frequency to obtain $\mathcal{A}^{loc}(\omega)$ and $\mathcal{A}^{aux}(\omega)$. Any reliable MaxEnt code should give similar results. The anomalous spectral weight is then extracted from the integral over wave vectors  of Eq.\eqref{Eq_Trick} 
\begin{equation}
\mathcal{A}^{an}(\omega) = \frac{1}{2}\, \left[ \mathcal{A}^{aux}(\omega) - \mathcal{A}^{loc}(\omega) - \mathcal{A}^{loc}(-\omega) \right].
\label{Eq_Trick_simplified}
\end{equation}
The imaginary parts of the local and anomalous Green functions are proportional to $\mathcal{A}^{loc}(\omega)$ and $\mathcal{A}^{an}(\omega)$ while their real parts are obtained using Kramers-Kr\"onig relations. The gap function then follows from~\cite{Poilblanc:2002}
\begin{equation}
\Delta(\omega)=\frac{2\omega\,\mathcal{F}(\omega)}{\mathcal{G}^{loc}(\omega)-\mathcal{G}^{loc\,*}(-\omega)}.
\label{Eq_Delta}
\end{equation}


\paragraph{Gap function} 

\begin{figure}[h!]
	\begin{center}
		\includegraphics[width=0.48\textwidth]{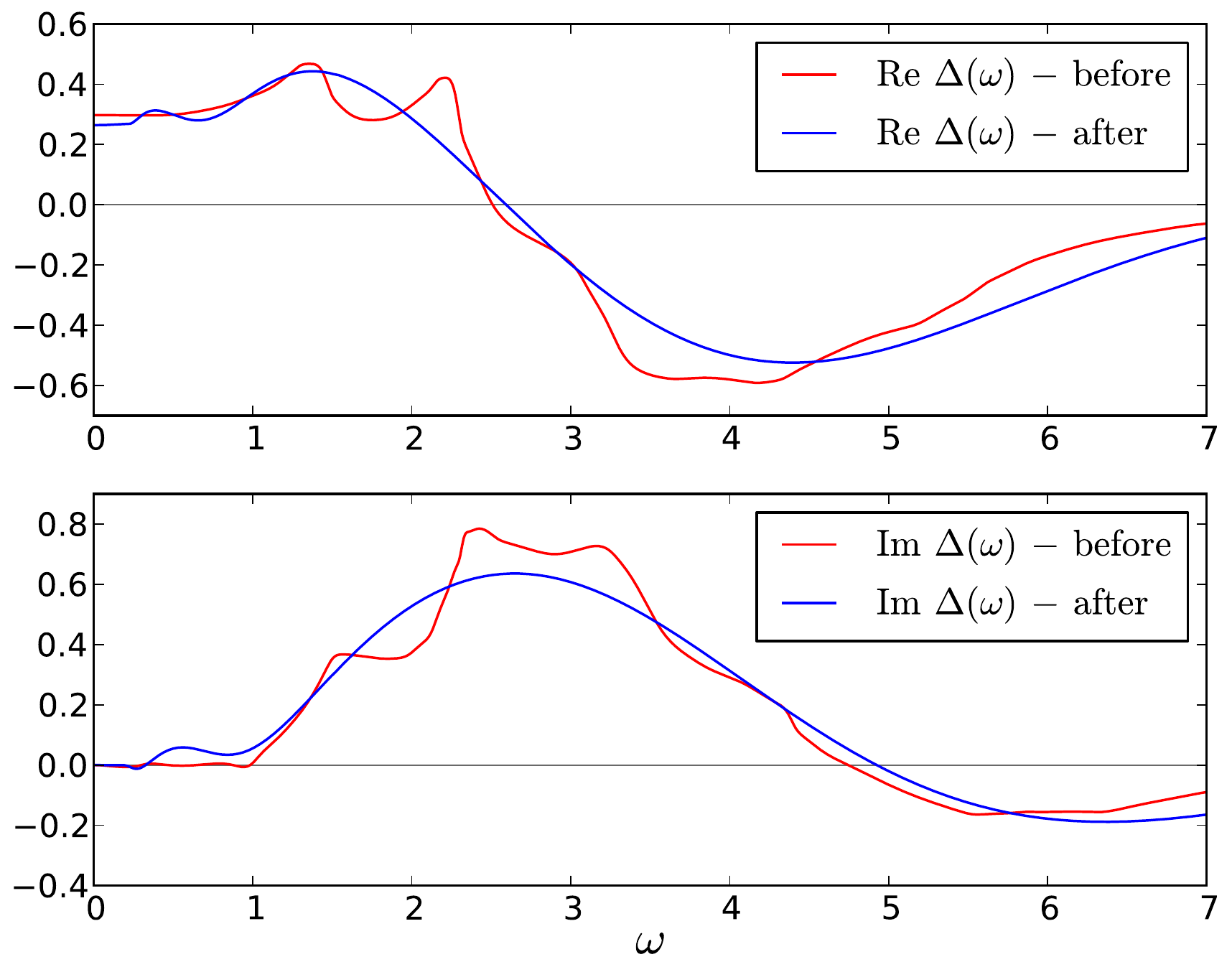}
		\caption{Red solid lines : spline interpolations of the real and imaginary parts of the gap function extracted from Ref.\cite{Scalapino:1966}, before performing the maximum-entropy analytic continuation. Blue solid lines : real and imaginary parts of the gap function obtained from Eq.\eqref{Eq_Delta}, after performing the maximum-entropy analytic continuation.}
		\label{Fig_Comparison_Delta}
	\end{center}
\end{figure}

One can now compare the original gap function of Fig.\ref{Fig_Interpolations_Delta} with the gap function Eq.\eqref{Eq_Delta} obtained from maximum-entropy analytic continuation. A direct comparison is shown in Fig.\ref{Fig_Comparison_Delta}.

Even though some of the details are lost in the analytic continuation, the main features are preserved, namely the gap function has a maximum and becomes negative at sufficiently high frequency before vanishing.


\paragraph{Gorkov function and Cumulative order parameter} 

\begin{figure}[h!]
	\begin{center}
		\includegraphics[width=0.48\textwidth]{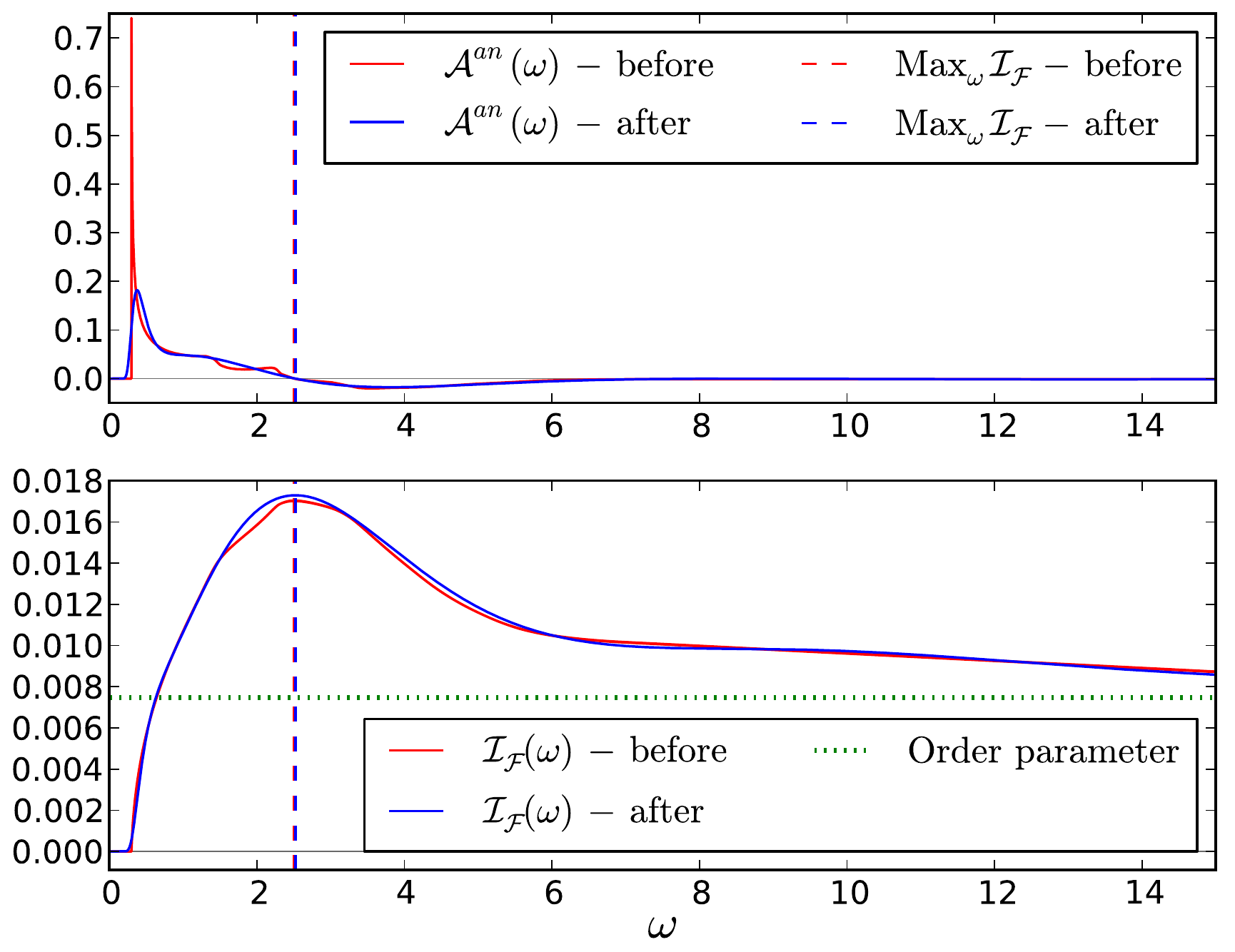}
		\caption{Red solid lines : anomalous spectral function and cumulative order parameter extracted from Eq.\eqref{Eq_DOS_SC}, before performing the maximum-entropy analytic continuation. Blue solid lines : anomalous spectral function and cumulative order parameter obtained from Eq.\eqref{Eq_Trick_simplified}, after performing the maximum-entropy analytic continuation. Dashed lines : frequency of the maximum of each cumulative order parameter. Green dotted line : value of the superconducting order parameter.}
		\label{Fig_Comparison_cumulative_order_parameter}
	\end{center}
\end{figure}

The top panel of Fig.\ref{Fig_Comparison_cumulative_order_parameter} compares the anomalous spectral function obtained from Eq.\eqref{Eq_DOS_SC} using the published $\Delta(\omega)$, with the same function extracted from Eq.\eqref{Eq_Trick_simplified} after analytic continuation of the Matsubara data. A negative component is clearly visible in both quantities above $2.5$ times the transverse phonon frequency. 

The cumulative order parameter, exhibited in the bottom panel of Fig.\ref{Fig_Comparison_cumulative_order_parameter}, contains the important information about the dynamics of pairing and it is clearly well preserved by the analytic continuation. At finite temperature, it is defined by:
\begin{equation}
\mathcal{I}_{\mathcal{F}}(\omega) = \int_{-\omega}^\omega \! \frac{\mathrm{d}\omega'}{2\pi} \, \mathcal{A}^{an}(\omega')\, f(-\omega')
\label{I_F}
\end{equation}
where $f(-\omega')=[1+e^{-\beta\omega'}]^{-1}$ is the Fermi-Dirac distribution. In the limit $\omega \rightarrow \infty$, we find that this quantity is just the order parameter $\mathcal{I}_{\mathcal{F}}(\infty) = \left\langle \hat{\mathcal{T}}_{\tau}\, \hat{c}_{i\uparrow}\, \hat{c}_{i\downarrow} \right\rangle_{\hat{\mathcal{H}}}$. At $T=0$, Eq.\eqref{I_F} reduces to the formulas in Refs.\cite{Kyung:2009,SenechalResilience:2013}.  In the same way that the dependence of $\left\langle \hat{\mathcal{T}}_{\tau}\, \hat{c}_{i\uparrow}\, \hat{c}_{j\downarrow} \right\rangle_{\hat{\mathcal{H}}}$ on distance $i-j$ tells us about the superconducting coherence length, the frequency dependence at the same position informs us on the dynamics.

As can be seen form Fig.\ref{Fig_Comparison_cumulative_order_parameter}, the frequencies larger than the gap contribute positively to the order parameter whose asymptotic value is represented by the horizontal dotted line. All frequencies up to the maximum, around $2.5$ times the transverse phonon frequency, contribute to pairing. After that, the effect of the Coulomb repulsion represented by $\mu^*$ in the Eliashberg equation are depairing. The position of the peak (vertical dashed line) and the region where depairing occurs are well preserved by the analytic continuation.

\paragraph{Auxiliary Green function for Analytic Continuation for General Multi-band Singlet or Triplet Pairing} 

Define,
\begin{equation}
\mathcal{G}_{\gamma}(\vec{k},\tau) = -\left\langle \hat{\mathcal{T}}_{\tau}\, \hat{c}_{\vec{k}\gamma}(\tau) \, \hat{c}^\dagger_{\vec{k}\gamma}(0) \right\rangle_{\hat{\mathcal{H}}}
\label{Eq_Normal_Green}
\end{equation}
where $\gamma$ stands for both band and spin index. Again, define the mixed operator
\begin{equation}
\hat{a}_{\vec{k}\gamma\delta} = \hat{c}_{\vec{k}\gamma} + \hat{c}^\dagger_{-\vec{k}\delta},
\label{Eq_mixed_operator-agd}
\end{equation}
and the auxiliary Green function 
\begin{equation}
\mathcal{G}^{aux1}_{\gamma\delta}(\vec{k},\tau) = -\left\langle \hat{\mathcal{T}}_{\tau}\, \hat{a}_{\vec{k}\gamma\delta}(\tau)\, \hat{a}^\dagger_{\vec{k}\gamma\delta}(0) \right\rangle_{\hat{\mathcal{H}}}.
\label{Eq_Total_Green_Function}
\end{equation}
The Lehmann representation allows one to check that this Green function has a positive spectral weight and can thus be analytically continued using standard Maximum Entropy methods. 

From the imaginary-time definition, one finds
\begin{widetext}
\begin{eqnarray}
\mathcal{G}^{aux\,1}_{\gamma\delta}(\vec{k},\tau) & = & \mathcal{G}_{\gamma} (\vec{k},\tau) - \mathcal{G}_{\delta} (-\vec{k},-\tau)  -\left\langle \hat{\mathcal{T}}_{\tau}\, \hat{c}_{\vec{k}\gamma}(\tau) \, \hat{c}_{-\vec{k}\delta}(0) \right\rangle_{\hat{\mathcal{H}}} -\left\langle \hat{\mathcal{T}}_{\tau}\, \hat{c}^{\dagger}_{-\vec{k}\delta}(\tau) \, \hat{c}^{\dagger}_{\vec{k}\gamma}(0) \right\rangle_{\hat{\mathcal{H}}}.
\label{Gaux1}
\end{eqnarray}
The Lehmann representation of each of the last two terms,
\begin{eqnarray}
-\left\langle \hat{\mathcal{T}}_{\tau}\, \hat{c}_{\vec{k}\gamma}(\tau) \, \hat{c}_{-\vec{k}\delta}(0) \right\rangle_{\hat{\mathcal{H}}} &=&-\frac{1}{\mathcal{Z}}\sum_{mn}e^{-\beta H_m}\langle m |e^{ H_m \tau} \hat{c}_{\vec{k}\gamma}e^{- H_n \tau}|n\rangle \langle n | \hat{c}_{-\vec{k}\delta}|m\rangle \\
-\left\langle \hat{\mathcal{T}}_{\tau}\, \hat{c}^{\dagger}_{-\vec{k}\delta}(\tau) \, \hat{c}^{\dagger}_{\vec{k}\gamma}(0) \right\rangle_{\hat{\mathcal{H}}}
&=& -\frac{1}{\mathcal{Z}}\sum_{mn}e^{-\beta H_m}\langle m |e^{H_m \tau} \hat{c}^{\dagger}_{-\vec{k}\delta}e^{- H_n \tau}|n\rangle \langle n | \hat{c}^{\dagger}_{\vec{k}\gamma}|m\rangle,
\end{eqnarray}
where $\mathcal{Z}$ is the grand partition function, and the equality
$
\lbrack\langle m | \hat{c}_{\vec{k}\gamma}|n\rangle \langle n | \hat{c}_{-\vec{k}\delta}|m\rangle\rbrack^*=
\langle m | \hat{c}^{\dagger}_{-\vec{k}\delta}|n\rangle \langle n | \hat{c}^{\dagger}_{\vec{k}\gamma}|m\rangle
$
show that in general the spectral weights of the two anomalous functions are complex conjugate of each other.  
Using the spectral representation of the anomalous spectral function,
\begin{equation}
\mathcal{F}_{\gamma\delta}(\vec{k},i\omega_n) = -\int_0^\beta \! \mathrm{d}\tau\, e^{i\omega_n\tau} \left\langle \hat{\mathcal{T}}_{\tau}\, \hat{c}_{\vec{k}\gamma}(\tau) \, \hat{c}_{-\vec{k}\delta}(0) \right\rangle_{\hat{\mathcal{H}}}=\int\!\frac{\mathrm{d}\omega}{2\pi}\,\frac{\mathcal{A}^{an}_{\gamma\delta}(\vec{k},\omega)}{i\omega_n-\omega},
\label{F}
\end{equation}
Eq.\eqref{Gaux1} can then be written as follows in Matsubara frequencies
\begin{eqnarray}
\mathcal{G}^{aux\,1}_{\gamma\delta}(\vec{k},i\omega_n) & = & \mathcal{G}_{\gamma} (\vec{k},i\omega_n) - \mathcal{G}_{\delta} (-\vec{k},-i\omega_n) +\int\!\frac{\mathrm{d}\omega}{2\pi}\,\frac{\mathcal{A}^{an}_{\gamma\delta}(\vec{k},\omega)+\mathcal{A}^{an\,*}_{\gamma\delta}(\vec{k},\omega)}{i\omega_n-\omega}.
\label{Gaux1Matsubara}
\end{eqnarray}
In practice, the auxiliary Green function in Matsubara frequency, that we need to analytically continue, would be computed using the results of Matsubara frequency calculations as follows:
\begin{eqnarray}
\mathcal{G}^{aux\,1}_{\gamma\delta}(\vec{k},i\omega_n) & = & \mathcal{G}_{\gamma} (\vec{k},i\omega_n) - \mathcal{G}_{\delta} (-\vec{k},-i\omega_n) +\mathcal{F}_{\gamma\delta}(\vec{k},i\omega_n)+[\mathcal{F}_{\gamma\delta}(\vec{k},-i\omega_n)]^*.
\end{eqnarray}
Eq.\eqref{Gaux1} can also be used instead if the analytic continuation is performed directly from imaginary time. 
\end{widetext}

To find the missing information, we define, by analogy with Eq.\eqref{Eq_mixed_operator-agd}, a mixed operator
\begin{equation}
\hat{b}_{\vec{k}\gamma\delta} = \hat{c}_{\vec{k}\gamma} + i\,\hat{c}^\dagger_{-\vec{k}\delta},
\label{Eq_mixed_operator_b}
\end{equation}
and a corresponding second auxiliary Green function with positive spectral weight 
\begin{equation}
\mathcal{G}^{aux\,2}_{\gamma\delta}(\vec{k},\tau) = -\left\langle \hat{\mathcal{T}}_{\tau}\, \hat{b}_{\vec{k}\gamma\delta}(\tau)\, \hat{b}^\dagger_{\vec{k}\gamma\delta}(0) \right\rangle_{\hat{\mathcal{H}}}.
\label{Eq_Total_Green_Function-2}
\end{equation}
Then, by analogy with Eq.\eqref{Gaux1Matsubara}, we find, 
\begin{widetext}
\begin{eqnarray}
\mathcal{G}^{aux\,2}_{\gamma\delta}(\vec{k},i\omega_n) & = & \mathcal{G}_{\gamma} (\vec{k},i\omega_n) - \mathcal{G}_{\delta} (-\vec{k},-i\omega_n) -i\int\!\frac{\mathrm{d}\omega}{2\pi}\,\frac{\mathcal{A}^{an}_{\gamma\delta}(\vec{k},\omega)-\mathcal{A}^{an\,*}_{\gamma\delta}(\vec{k},\omega)}{i\omega_n-\omega}.
\label{Gaux2Matsubara}
\end{eqnarray}
The analytic continuation of the two auxiliary Green functions gives the corresponding real and positive spectral weights, which are formally given by
\begin{eqnarray}
\mathcal{A}_{\gamma\delta}^{aux\,1}(\vec{k},\omega)&=& \mathcal{A}_\gamma(\vec{k},\omega)+\mathcal{A}_\delta(-\vec{k},-\omega)+\mathcal{A}^{an}_{\gamma\delta}(\vec{k},\omega)+\mathcal{A}^{an\,*}_{\gamma\delta}(\vec{k},\omega)\\
\mathcal{A}_{\gamma\delta}^{aux\,2}(\vec{k},\omega)&=& \mathcal{A}_\gamma(\vec{k},\omega)+\mathcal{A}_\delta(-\vec{k},-\omega)-i(\mathcal{A}^{an}_{\gamma\delta}(\vec{k},\omega)-\mathcal{A}^{an\,*}_{\gamma\delta}(\vec{k},\omega)).
\end{eqnarray}
The normal spectral functions are also real and positive. So, from the above, one can easily extract the needed spectral weight $\mathcal{A}^{an}_{\gamma\delta}(\vec{k},\omega)$ via
\begin{equation}
\mathcal{A}^{an}_{\gamma\delta}(\vec{k},\omega) = \frac{1}{2} \left[ \mathcal{A}_{\gamma\delta}^{aux\,1}(\vec{k},\omega) + i\,\mathcal{A}_{\gamma\delta}^{aux\,2}(\vec{k},\omega) -(1+i)\left(\mathcal{A}_\gamma(\vec{k},\omega)+\mathcal{A}_\delta(-\vec{k},-\omega)\right)\right].
\end{equation}
\end{widetext}
Recalling that $\gamma$ and $\delta$ contain both spin and band index, this is valid for singlet or triplet pairing and for the multi-orbital case, with or without inversion or time-reversal symmetry.

\paragraph{Conclusion} The above results open the way to the systematic exploration of the pairing mechanism in numerical calculations for unconventional superconductors. Similar auxiliary functions can be defined to find, with Maximum Entropy methods, spectral functions of transport coefficients, such as the thermoelectric power,  that do not have positive spectral weight. 

\paragraph{Acknowledgments}
We are grateful to M. Charlebois, D. S\'en\'echal and S. Verret for useful discussions. This work has been supported by the Natural Sciences and Engineering Research Council of Canada (NSERC), and by the Tier I Canada Research Chair Program (A.-M.S.T.).


\end{document}